\documentclass[graybox]{svmult}

\usepackage{mathptmx}       
\usepackage{helvet}         
\usepackage{courier}        
\usepackage{type1cm}        
\usepackage{graphicx}        
\usepackage{multicol}        
\usepackage[bottom]{footmisc}

\usepackage{url}
\usepackage{amsfonts}
\usepackage{amssymb}
\usepackage[utf8]{inputenc}
\usepackage{tabularx}
\usepackage{mdwlist}
\usepackage{afterpage}

\newcommand{\comment}[1]{}

\newtheorem{notation}{Notation}[section]

\newcounter{fnc}

\usepackage[olditem,oldenum]{paralist}

\newenvironment{paraenum}{%
  \begin{inparaenum}[i)]}
  {\end{inparaenum}}

\usepackage{xspace}
\usepackage{extarrows}
\usepackage{prooftree}
\usepackage{stmaryrd}

\usepackage{color}

\definecolor{RuleColor}{rgb}{0.7,0.7,0.7}

\newcommand{\vertrel}[0]{\mathrel{\vert}}

\newcommand{\vv}[1]{\overline{#1}}

\newcommand{\Trm}[1]{\textrm{#1}}

\def\bottom{\mathop{\perp}}

\makeatletter
\def\sfrac#1/#2{\leavevmode\kern.1em
        \raise.5ex\hbox{\the\scriptfont\z@ #1}\kern-.1em
        /\kern-.15em\lower.25ex\hbox{\the\scriptfont\z@ #2}}
\makeatother

\newcommand{\ra}[0]{\rightarrow}

\newcommand{\da}[0]{\downarrow}

\newcommand{\uv}[1]{\underline{#1}}
\newcommand{\ov}[1]{\overline{#1}}

\newcommand{\CC}{\textrm{CC}\xspace}
\newcommand{\CIC}{\textrm{CIC}\xspace}

\newcommand{\CCIC}{\textrm{CCIC}\xspace}
\newcommand{\CCNAT}{{\ensuremath{\textrm{CC}_\sN}}\xspace}

\newcommand{\Prop}   {\ensuremath{\star}\xspace}
\newcommand{\Type}   {\ensuremath{{\Box}}\xspace}
\newcommand{\Extern} {\ensuremath{\triangle}\xspace}

\DeclareMathOperator{\Ind}{Ind}
\DeclareMathOperator{\Elim}{Elim}

\DeclareMathOperator{\FV}{FV}

\DeclareMathOperator{\dom}{dom}

\newcommand{\fosort}[1]{\ensuremath{\mathbf{#1}}}
\newcommand{\fosymb}[1]{\ensuremath{\mathbf{#1}}}

\newcommand{\nat}{\fosort{nat}}
\newcommand{\lst}{\fosort{list}}

\newcommand{\NO}{\fosymb{0}}
\newcommand{\NS}{\fosymb{S}}

\newcommand{\NP}{\mathop{\dot{+}}}

\newcommand{\cons}{\fosymb{cons}}
\newcommand{\nil}{\fosymb{nil}}

\newcommand{\app}{\fosymb{app}}
\newcommand{\car}{\fosymb{car}}

\newcommand{\word}{\fosort{word}}
\newcommand{\mv}{\epsilon}
\newcommand{\chr}{\fosort{char}}

\newcommand{\ccieq}[0]{\mathop{\dot{=}}}

\newcommand{\ElimTS}[5]{\Elim^\cS({#1} : {#2} \, [{#3}] \ra {#4})\{{#5}\}}

\newcommand{\ElimT}[5]{\Elim({#1} : {#2} \, [{#3}] \ra {#4})\{{#5}\}}

\newcommand{\ElimnatF}[0]{\Elim\!\sN}
\newcommand{\ElimwordF}[0]{\Elim\!\sW}
\newcommand{\ElimlistF}[0]{\Elim\!\sL}
\newcommand{\Elimnat}[2]{\ElimnatF({#1})\{{#2}\}}
\newcommand{\Elimlist}[2]{\Elim\!\sL({#1})\{{#2}\}}
\newcommand{\Elimword}[2]{\ElimwordF({#1})\{{#2}\}}

\newcommand{\ProdBaseT}[2] {\forall {#1} .\, {#2}}
\newcommand{\AbsBaseT}[2]  {\lambda {#1} .\, {#2}}

\newcommand{\ProdSur}[1]{({#1})}
\newcommand{\AbsSur} [1]{[{#1}]}

\newcommand{\ProdT}[2] {\ProdBaseT {\ProdSur{#1}} {#2}}
\newcommand{\AbsT}[2]  {\AbsBaseT  {\AbsSur {#1}} {#2}}

\newcommand{\ProdTT}[3]%
  {\ProdBaseT {\ProdSur{#1} \ProdSur{#2}} {#3}}

\newcommand{\AbsTT}[3]%
  {\AbsBaseT  {\AbsSur {#1} \AbsSur{#2}}  {#3}}

\newcommand{\ProdTTT}[4]%
  {\ProdBaseT {\ProdSur{#1} \ProdSur{#2} \ProdSur{#3}} {#4}}

\newcommand{\AbsTTT}[4]%
  {\AbsBaseT  {\AbsSur {#1} \AbsSur {#2} \AbsSur{#3}}  {#4}}

\newcommand{\AbsTTTT}[5]%
  {\AbsBaseT  {\AbsSur {#1} \AbsSur {#2} \AbsSur{#3} \AbsSur{#4}}  {#5}}

\newcommand{\AbsTTTTT}[6]%
  {\AbsBaseT  {\AbsSur {#1} \AbsSur {#2} \AbsSur{#3} \AbsSur{#4} \AbsSur{#5}}  {#6}}

\DeclareMathOperator{\Eq}{Eq}

\DeclareMathOperator{\Class}{Class}

\DeclareMathOperator{\cnvop}{\sim}
\DeclareMathOperator{\cnvblop}{\approx}

\newcommand{\cnv}[1]{\cnvop_{#1}}
\newcommand{\cnvbl}[1]{\cnvblop_{#1}}

\newcommand{\nf}[2]{{#1}\da_{#2}}

\newcommand{\Ar}{\textrm{r}}
\newcommand{\Au}{\textrm{u}}

\DeclareMathOperator{\Alg}{\cA}
\DeclareMathOperator{\Algt}{{\cA}lg}

\newcommand{\rulename}[1]{\textsc{[#1]}}

\newlength  {\CInferRuleWidth}
\newlength  {\CInferRuleWhenWidth}
\newsavebox {\CInferRuleBox}

\newcommand{\InferRule}[3]{%
\prooftree
  \: \mbox{%
      $\begin{array}{@{}c@{}}
        #2
      \end{array}$}%
  \:
\justifies
  \: {#3} \:
\using
  \mbox{\small \rulename{#1}}
\endprooftree}

\newcommand{\NInferRule}[2]{%
\prooftree
  \: \mbox{%
      $\begin{array}{@{}c@{}}
        #1
      \end{array}$}%
  \:
\justifies
  \: {#2} \:
\endprooftree}

\newcommand{\rw}[1]{\mathop{\xrightarrow{#1}}}

\newcommand{\rwequiv}[1]{\mathop{\xleftrightarrow{#1}_*}}
\newcommand{\rwrefltr}[1]{\mathop{\xrightarrow{#1}_*}}

\newcommand{\cA}[0]{\mathcal{A}}

\newcommand{\cD}[0]{\mathcal{D}}

\newcommand{\cK}[0]{\mathcal{K}}
\newcommand{\cL}[0]{\mathcal{L}}
\newcommand{\cM}[0]{\mathcal{M}}

\newcommand{\cO}[0]{\mathcal{O}}
\newcommand{\cP}[0]{\mathcal{P}}

\newcommand{\cR}[0]{\mathcal{R}}
\newcommand{\cS}[0]{\mathcal{S}}
\newcommand{\cT}[0]{\mathcal{T}}

\newcommand{\cV}[0]{\mathcal{V}}
\newcommand{\cW}[0]{\mathcal{W}}
\newcommand{\cX}[0]{\mathcal{X}}
\newcommand{\cY}[0]{\mathcal{Y}}

\newcommand{\sL}[0]{\mathbb{L}}

\newcommand{\sN}[0]{\mathbb{N}}

\newcommand{\sW}[0]{\mathbb{W}}

\begin{document}

\title*{From Formal Proofs to Mathematical Proofs:\\
 A Safe, Incremental Way for Building in First-order Decision Procedures}
\titlerunning{From Formal Proofs to Mathematical Proofs}
\author{Fr\'ederic Blanqui and Jean-Pierre Jouannaud and Pierre-Yves Strub}
\authorrunning{F. Blanqui and J.-P. Jouannaud and P.-Y. Strub}
\institute{Fr\'ederic Blanqui\at
INRIA \& LORIA, Campus Scientifique, BP 239, 54506 Vandoeuvre-l\`es-Nancy Cedex, France, \email{blanqui@loria.fr}
\and Jean-Pierre Jouannaud \and Pierre-Yves Strub\at
LIX, UMR 7161, Project INRIA TypiCal,
\'Ecole Polytechnique, 91128 Palaiseau, France, \email{jouannaud,strub@lix.polytechnique.fr}}
%
%
\maketitle
\vspace*{-2.5cm}

\abstract{We investigate here a new version of the Calculus of Inductive
Constructions (CIC) on which the proof assistant Coq is based: the
Calculus of Congruent Inductive Constructions, which truly extends CIC
by building in arbitrary first-order decision procedures: deduction is
still in charge of the CIC kernel, while computation is outsourced to
dedicated first-order decision procedures that can be taken from the
shelves provided they deliver a proof certificate. The soundness of
the whole system becomes an incremental property following from the
soundness of the certificate checkers and that of the kernel. A
detailed example shows that the resulting style of proofs becomes
closer to that of the working mathematician.}
\vspace*{-.5cm}

\section{Introduction}
\label{s:intro}

Proof assistants based on the Curry-Howard isomorphism such as Coq
\cite{coqv80} allow to build the proof of a given proposition by
applying appropriate proof tactics available from existing libraries
or that can otherwise be developed for achieving a specific task.
These tactics generate a proof term that can be checked with respect
to the rules of logic. The proof-checker, also called the {\em
kernel} of the proof assistant, implements the deduction rules of the
logic on top of a term manipulation layer. In this model, the
mathematical correctness of a proof development relies entirely on the
kernel. Trusting the kernel is therefore vital.

The (intuitionist) logic on which Coq is based is the Calculus of
Constructions (CC) of Coquand and Huet \cite{coquand:ic88}, an
impredicative type theory incorporating polymorphism, dependent types
and type constructors. Unlike logics without dependent types, CC
enjoys a powerful type-checking rule, called {\em conversion}, which
incorporates computations within deductions, making decidability of
type-checking a non-trivial property of the calculus.

In CC, computation reduces to (pure) functional evaluation in the
underlying lambda calculus. The notion of computation is richer in the
Calculus of Inductive Constructions of Coquand and Paulin (CIC),
obtained from CC by adding inductive types and the corresponding rules
for higher-order primitive recursion \cite{coquand89}. The recent
versions of Coq are based on a slight generalization of this calculus
\cite{gimenez:icalp98}. Still, such a simple function as {\em
reverse} of a {\em dependent list} cannot be defined in CIC as one
would expect, because $(reverse ~l :: l')$ and $(reverse~l') ::
(reverse~l)$, assuming $::$ is list concatenation, have
non-convertible types $list(n+m)$ and $list(m+n)$, assuming
$(reverse~l)$ has for type the type of its argument $l$. This is so
because the usual definition of $+$ by induction on one of its
arguments does not reduce the proof of $m+n=n+m$ to a computation.

We do believe that scaling up the proof development process requires
being able to mimic the mathematician when replacing the proof of a
proposition P by the proof of an equivalent proposition P' obtained
from P thanks to possibly complex calculations in which {\em easy
steps} are hidden away. It is our program to make this view a reality.

A way to incorporate decision procedures to Coq is by developing a
tactic and then use a reflexion technique to omit checking the proof
term being built by proving the decision procedure itself. But the
soundness of the entire mechanism cannot be guaranteed in general
\cite{corbineau:thesis}. Further, this does not answer the question of
hiding easy steps away.

A first attempt towards our goal is the Calculus of Algebraic
Constructions (CAC), obtained by adding to CC user-defined
computations as rewrite rules
{}\cite{blanqui99rta,blanqui:mscs05}. Although conceptually quite
powerful since CAC captures CIC \cite{blanqui:fi05}, this paradigm
does not yet fulfill all needs. In particular, the user needs to hide
away the easy steps by himself, that is by giving the necessary
rewrite rules and by verifying that they satisfy the assumptions of
the {\em general schema} \cite{blanqui99rta,blanqui:mscs05}.

The proof assistant PVS uses a potentially stronger paradigm than Coq
by combining its deduction mechanism with a notion of computation
based on the powerful Shostak's method for combining decision
procedures \cite{shostak79}, a framework dubbed {\em little proof
engines} by Shankar \cite{shankar:lics02}. Indeed, the little engines
of proof hide away the easy computational steps, without any user
assistance. Unfortunately, proof-checking is not decidable in
PVS. Further, since the little engines of proofs involve complex
coding, as well as Shostak's algorithm itself, one can only {\em
believe} a PVS proof, while one can {\em check} and {\em trust} a Coq
proof.

Two steps in the direction of integrating decision procedures into CC
are Stehr's Open Calculus of Constructions (OCC) \cite{stehr:fi07} and
Oury's Extensional Calculus of Constructions (ECC)
{}\cite{oury:tphol}. Implemented in Maude, OCC allows for the use of
an arbitrary equational theory in conversion. ECC can be seen as a
particular case of OCC in which all provable equalities can be used in
conversion, which can also be achieved by adding the extensionality
and Streicher's axioms to CC \cite{streicher93phd}, hence the name of
this calculus. Unfortunately, strong normalization and decidability of
type checking are then lost\comment{in ECC (and OCC)}, which shows
that we should seek for more restrictive extensions.

In a preliminary work, we designed a new, quite restrictive framework,
the Calculus of Congruent Constructions (CCC), which incorporates the
congruence closure algorithm in CC's conversion \cite{ccc-draft},
while preserving the good properties of CC, including the decidability
of type checking. In \cite{blanqui07csl}, we have described \CCNAT, in
which the decision procedure was Presburger arithmetic and strong
elimination ruled out. The present work is a continuation of the
latter.

{\bf Theoretical contribution.} Our main theoretical contribution is
the definition and the meta-theoretical investigation of the Calculus
of Congruent Inductive Constructions (CCIC), which incorporates
arbitrary {\em first-order theories} for which entailment is decidable
into deductions via an abstract conversion rule of the calculus. A
major technical innovation of this work lies in the computation
mechanism: goals are sent to the decision procedure together with the
set of user hypotheses available from the current context. Our main
result shows that this extension of CIC does not compromise its
properties: confluence, strong normalization, coherence and
decidability of proof-checking are all preserved.

Unlike previous calculi, the difficulty with CCIC is not strong
normalization, for which we have reused the strong normalization proof
of CAC \cite{blanqui:mscs05}. A major difficulty was a traditional
step towards subject-reduction: compatibility of conversion with
products. Decidability of type checking required restricting
conversions below recursors \cite{strub:thesis}.

{\bf Practical contribution.} We give several examples showing the
usefulness of this new calculus, in particular for using dependent
types such as dependent lists, which has been an important weakness of
Coq until now. Further studies are needed to explore other potential
applications, to match inductive definition-by-case modulo theories of
constructors-destructors, another very different weakness of Coq. A
detailed example shows that the resulting style of proofs becomes
closer to that of the working mathematician.

{\bf Methodological contribution.} The safety of proof assistants is
based on their kernel. In the early days of Coq, the safety of its
kernel relied on its small size and its clear structure reflecting the
inference rules of the intuitionist type theory, CC, on which it was
based. The slogan was that of a {\em readable kernel}. Moving later to
CIC allowed to ease the specification tasks, making the system very
popular among proof developers, but resulted in a more complex kernel
that can now hardly be read except by a few specialists. The slogan
changed to a {\em provable kernel}, and indeed one version of it was
once proved with an earlier version (using strong normalization as an
assumption), and a new safe kernel extracted from that proof.

Of course, there has been many changes in the kernel since then, and
its correctness proof was not maintained. This is a first weakness
with the {\em readable kernel} paradigm: it does not resist
changes. There is a second which relates directly to CCIC: there is no
guarantee that a decision procedure taken from the shelf implements
correctly the complex mathematical theorem on which it is based, since
carrying out such a proof may require an entire PhD work. Therefore,
these procedures {\em cannot} be part of the kernel.\comment{A major reason
why no implementation of CAC was ever released, is indeed that making
type-checking efficient would require compiling the user-defined
rules, a complex task resulting in a kernel too large to be trusted
anymore or even proved.}

Our solution to these problems is a new shift of paradigm to that of
an {\em incremental kernel}. The calculus on which a proof assistant
is based should come in two parts: a stable calculus implementing
deduction, CIC in our case, which should satisfy the {\em readable} or
{\em provable kernel} paradigm; a collection of independent decision
procedures implementing computations, that produce checkable proof
certificates. The certificate checker should of course itself satisfy
the {\em readable} or {\em provable code} paradigm. Note that a Coq
proof is a particular case of a checkable certificate.

This paradigm has many advantages. First, it allows for a modular,
cooperative development of the system, by separating the development
of the kernel from that of the decision procedures. Second, it allows
for an {\em unsafe mode} in case a decision procedure is used that
does not have a certificate generator yet. Third, it allows to better
trace errors in case the system rejects a proof, by using decision
procedures that output {\em explanations} when they fail. Last, it
allows the user \comment{herself} to use any decision procedure she
needs by simply hooking it to the system, possibly in unsafe mode.

This incremental schema is quite flexible, assuming that decision
procedures come one by one. However, even so, they are not
independent, they must be combined. Combining first-order decision
procedures is not a new problem, it was considered in the early 80's
by Nelson and Oppen on the one hand, by Shostak on the other hand,
and has generated much work since then. There are several
possibilities to build in this mechanism: in the kernel, via a
certificate generator and checker again, or by reflection. This
design decision has not been made yet.

\comment{We assume some familiarity with typed lambda calculi
\cite{barendregt:book92} and the Calculus of Inductive Constructions
\cite{werner:thesis,blanqui:fi05}.}

\vspace*{-0.5cm}
\section{Congruent Inductive Constructions}
\label{s:cic}

The Calculus of Congruent Inductive Constructions (CCIC) is an
extension of CIC which embeds in its conversion rule the validity
entailment of a fixed first order theory. First, we recall the basics
of CIC before to introduce parametric multi-sorted algebras and then
embed these first-order algebras into CIC. We are then able to define
our calculus relative to a specific congruence that is defined last.
For simplicity, we will only consider here the particular case of
parametric lists and that of the natural numbers equipped with
Presburger arithmetic. This simple case allows us to build lists of
natural numbers, as well as lists of lists of natural numbers, and so
on. It indeed has the complexity of the whole calculus, which is not
at all the case when natural numbers only are considered as in
\cite{blanqui07csl}.


\vspace*{-0.5cm}
\subsection{Calculus of Inductive Constructions}
\label{ss:cic}

{\bf Terms.} We start our presentation by first describing the terms
of CIC.

{}\CIC uses two {\em sorts}: \Prop (or Prop, or {\em object level
  universe}), \Type (or Type, or {\em predicate level universe}) and
\Extern. We denote $\{ \Prop, \Type, \Extern \}$, the set of \CIC
sorts, by $\cS$.

Following the presentation of {\em Pure Type Systems} (PTS)
\cite{geuvers:jfp89}, we use two classes of variables: $\cX^\Prop$ and
$\cX^\Type$ are countably infinite sets of {\em term variables} and
{\em predicate variables} such that $\cX^\Prop$ and $\cX^\Type$ are
disjoint. We write $\cX$ for $\cX^\Prop \cup\cX^\Type$.

We shall use
  $\ov{u}$ for a list $(u_1,\ldots,u_n)$,
  $s$            for a sort in $\cS$,
  $x, y, \ldots$ for variables in $\cX^\Prop$,
  $X, Y, \ldots$ for variables in $\cX^\Type$.

\begin{definition}[Pseudo-terms]
\label{def:cic-terms}
The algebra $\cL$ of {\em pseudo-terms} of \CIC is defined by:

\noindent\centering
  \begin{tabular}{r@{$\;$}c@{$\;$}l}
    $t, u, T, U, \ldots$ & $:=$
      & $s \in \cS \vertrel
         x \in \cX \vertrel
         \ProdT {x : T}{t} \vertrel
         \AbsT  {x : T}{t}$ \\[.3em]
     & $\vertrel$ & $t \, u \vertrel \Ind(X : t)\{ \vv{T_i} \}
     \vertrel  t^{[n]} \vertrel \ElimT{t}{T}{\vv{u_i}}{U}{\vv{w_j}}$
  \end{tabular}
\end{definition}

The notion of free variables is as usual - the binders being
$\lambda$, $\forall$ and $\Ind$ (in $\Ind(X : t)\{\vv{T_i}\}$, $X$ is
bound in the $T_i$'s). We write $\FV(t)$ for the set of free variables
of $t$. We say that $t$ is closed if $\FV(t) =
\emptyset$. A variable $x$ {\em freely occurs} in $t$ if $x \in
\FV(t)$.\smallskip


{\bf Inductive types.} The novelty of \CIC was to introduce inductive
types, denoted by $I =\Ind(X : T)\{ \vv{C_i} \}$ where the
$\vv{C_i}$'s describe the types of the {\em constructors} of $I$, and
$T$ the type (or {\em arity}) of $I$ which must be of the form
$\ProdT{\vv{x_i : T_i}}{\Prop}$.  The $k$-th constructor of the
inductive type $I$, of type $C_k \{ X\mapsto I \}$, will be denoted by
$I^{[k]}$.

As an easy first example, we define natural numbers:
{\small${\nat} := \Ind(X : \Prop)\{X, X\ra X\}$}.
We shall use $\NO$ and $\NS$ as constructors for natural numbers, of
respective types $\nat$ and $\nat\ra\nat$, obtained by replacing $X$
by $\nat$ in the above two expressions $X$ and $X\ra X$.  Elimination
rules for $\nat$ are as follows:

\begin{center}\small
\begin{tabular}{@{}l@{$\:\rw{\iota}\:$}l@{}}
  $\Elimnat{\NO, Q}{v_\NO, v_\NS}$ & $v_\NO$ \\[.3em]
  $\Elimnat{\NS \, x ,Q}{v_\NO, v_\NS}$ & $ v_\NS \, x \,
(\Elimnat {x, Q} {v_\NO, v_\NS})$ with $Q : \nat \ra s,~ \in\cS$.
\end{tabular}
\end{center}

\noindent
Similarly, we now define parametric lists:
{\small${\lst} := \AbsT{T : \Prop}{\Ind(X : \Prop)\{X, T\ra X \ra X\}}$}.
We shall use $\nil$ and $\cons$ as constructors for parametrized
lists, of respective types $\ProdT{T : \Prop}{\lst(T)}$ and $\ProdT{T
: \Prop}{T \ra \lst(T) \ra\lst(T)}$.\comment{, obtained by replacing $X$ by
$\lst(T)$ in $X$ and $T \ra X\ra X$, and then abstracting over the
type $T:\Prop$.} Elimination rules for $\lst$ are:

\begin{center}\small
\begin{tabular}{@{}l@{$\:\rw{\iota}\:$}l@{}}
  $\Elimlist{\nil, Q}{v_\nil, v_\cons}$ & $v_\nil$ \\[.3em]
  $\Elimlist{\cons \, x \, l, Q}{v_\nil, v_\cons}$ & $v_\cons \, x \, l\,
\Elimlist {l, Q} {v_\nil, v_\cons})$\\[.3em]
\end{tabular}
\end{center}

\noindent
Finally, we define dependent words over an alphabet $A$:
\[\small
\begin{array}{rcr}
\word &=& \Ind(X : \nat\ra \Prop)\{ {X \, \NO}, {A\ra X\ (\NS\, \NO)},
\ProdT{y,z:\nat}{X\,y \ra X\,z \ra X (y+z)}\}
\end{array}
\]
We shall use $\mv$, $\chr$ and $\app$ for its three constructors, of
respective types $\word \, \NO$, $A \ra \word\, (\NS\, \NO)$, and
$\ProdT{n,m:\nat}{\word\, n \ra \word\, m \ra \word\, (n+m)}$ obtained
as previously by replacing $X$ by $\word$ in the three expressions ${X
\, \NO}, {A\ra X\ (\NS\, \NO)},$ and $\ProdT{y,z:\nat}{X\,y \ra X\,z
\ra X (y+z)}$. Elimination rules for dependent words are:

\begin{center}\small
\begin{tabular}{@{}l@{$\:\rw{\iota}\:$}l@{}}
  $\Elimword{\mv, Q}{v_\mv, v_\chr, v_\app}$ & $v_\mv$ \\[.3em]
  $\Elimword{\chr\, x ,Q}{v_\mv, v_\chr, v_\app}$ & $ v_\chr\, x$\\[.3em]
  $\Elimword{\app\, n\, m\, l\, l', Q}{v_\mv, v_\chr, v_\app}$ & 
$v_\app \, n \, m\, l\, l'\,
(\Elimword{l, Q}{v_\mv, v_\chr, v_\app})$\\[.2em]
\multicolumn{2}{r}{$ (\Elimword{l', Q}{v_\mv, v_\chr, v_\app})$}\\[.2em]
\end{tabular}
\end{center}


{\bf Definitions by induction.} We can now define functions by
induction over natural numbers, lists or words. Since using the \CIC
syntax is a bit painful, we give only a quite simple example defining
append (written $@$) for lists of natural numbers, of type $\ProdT{T
: \Prop}{\lst(T) \ra \lst(T) \ra\lst(T)}$:

$\small\begin{array}{@{}l@{}}
@ := \AbsTT{l : \lst \, \nat}{l' : \lst \, \nat}{}
       {\ElimlistF
           (l, Q)
           \left\{l',\begin{array}{@{}l@{}}
              \AbsTT{x : \nat}{l'' : \lst \, \nat}{}\\
              \quad\AbsTT{l1 : \lst \, \nat}{l2 : \lst \, \nat}{}\\
              \quad\quad\AbsT{L : Q \, l1 \, l2}{\cons \, x \, L}\\
          \end{array}\right\}}
\end{array}$\\[1mm]

{\bf Strong and Weak reductions.} \CIC distinguishes {\em strong}
$\iota$-elimination when the type $Q$ of terms constructed by
induction is at predicate level, from weak $\iota$-elimination when
$Q$ is at object level.  Strong elimination is restricted to {\em
small} inductive types to ensure logical consistency
\cite{werner:thesis}.\smallskip

{\bf Typing judgments.} A {\em typing environment} $\Gamma$ is a
sequence of pairs $x_i : T_i$ made of a variable $x_i$ and a term
$T_i$ (we say that $\Gamma$ binds $x_i$ to the type $T_i$), such that
$\Gamma$ does not bind a variable twice.  The typing judgments are
classically written $\Gamma \vdash t : T$, meaning that the {\em well
formed term} $t$ is a proof of the proposition $T$ (has type $T$)
under the {\em well} {\em formed environment} $\Gamma$.  $x\Gamma$
will denote the type associated to $x$ in $\Gamma$, and we write
$\dom(\Gamma)$ for the domain of $\Gamma$ as well.

The typing rules of \CIC given in \ref{fig:cic-rules-ind} are made of
the typing rules for CC\comment{given at Figure
\ref{fig:cic-rules}} and the typing rules for inductive types, given
\comment{at Figure {}\ref{fig:cic-rules-ind}} for the particular case of $\nat$
and $\lst$.

\begin{figure}
\centering\small
\begin{tabular}{cc}
\begin{minipage}{6cm}\centering
\InferRule{Ax-1}{}{\vdash \Prop : \Type}
\hspace{1em}
\InferRule{Ax-2}{}{\vdash \Type : \Extern}\\[2mm]

\InferRule{Prod}
  {\Gamma \vdash T : s_T \quad \Gamma,[x : T] \vdash U : s_U}
  {\Gamma \vdash \ProdT{x : T}{U} : s_U}\\[2mm]

\InferRule{Abs}
  {\Gamma \vdash \ProdT{x : T}{U} : s \quad \Gamma, [x : T] \vdash u : U}
  {\Gamma \vdash \AbsT{x : T}{u} : \ProdT{x : T}{U}}\\[2mm]

\InferRule{App}
          {\Gamma \vdash t : \ProdT{x : U}{V} \quad
           \Gamma \vdash u : U}
          {\Gamma \vdash t\,u : V \{ x \mapsto u \}}
\end{minipage}
&\begin{minipage}{6cm}\centering
\InferRule{Weak}
  {\Gamma \vdash V : s      \quad
   \Gamma \vdash t : T      \quad
   s \in \{ \Prop, \Type \} \\
   x \in \cX^s - \dom(\Gamma)}
  {\Gamma, [x : V] \vdash t : T}\\[2mm]

\InferRule{Var}
  {x \in \dom(\Gamma) \cap \cX^{s_x} \quad \Gamma \vdash x\Gamma : s_x}
  {\Gamma \vdash x : x\Gamma}\\[2mm]

\InferRule{Conv}
  {\Gamma \vdash t : T \quad \Gamma \vdash T' : s' \quad T \rwequiv{\beta\iota} T'}
  {\Gamma \vdash t : T'}
\end{minipage}
\end{tabular}\\[2mm]
\comment{
\smallskip
\caption{\label{fig:cic-rules} \CIC typing rules (\CC rules)}
\end{figure}


\begin{figure}[h]
\centering\small}
\begin{tabular}{cc}
\begin{minipage}{5.5cm}\centering
\InferRule{Symb}
  {\vdash \tau_f : s \in \{ \Prop, \Type \}}
  {\vdash f:\tau_f}\\[2mm]

\InferRule{Elim}
  {\Gamma \vdash Q : \nat \ra s\in\{\Prop, \Type\}\\
   \Gamma \vdash n : \nat \quad
   \Gamma \vdash v_\NO : Q\,\NO \\
   \Gamma \vdash v_\NS : \ProdT{p : \nat}{Q\,p \ra Q\,(\NS\,p)}}
  {\Elimnat{n, Q}{v_\NO, v_\NS} : Q\,n}
\end{minipage}
&\begin{minipage}{6cm}\centering
\InferRule{Elim}
  {\Gamma \vdash T : \Prop \quad
   \Gamma \vdash p : \nat \quad
   \Gamma \vdash l : \lst \, T \, p \\[.5em]
   \Gamma \vdash Q : \ProdT{n : nat}{\lst\,T\,n \ra s\in\{\Prop, \Type\}} \\[.2em]
   \Gamma \vdash v_\nil : Q \, \NO \, (\nil \, T) \\[.5em]
   \Gamma \vdash v_\cons :
      \begin{array}{@{}l@{}}
        \ProdTTT{x : T}{n : \nat}{l : \lst\,T\,n}{}\\
          \quad {Q\,n\,l \ra Q\,(\NS\,n)(\cons\,T\,x\,n\,l)}
      \end{array}}
  {\Elimlist{l, Q}{v_\NO, v_\NS} : Q\,p\,l}
\end{minipage}
\end{tabular}
\smallskip
\caption{\label{fig:cic-rules-ind}\CIC typing rules for $\nat$ and $\lst$}
\end{figure}

We did not give the general typing elimination rule for arbitrary
inductive types, which is quite complicated. Instead, we gave the
elimination rules obtained for our three inductive types $\nat, \lst$
and $\word$.  We refer to \cite{paulin93tlca,werner:thesis} for the
general case, and for the precise typing rule of $\Elim\!\sW$.

\comment{It can be easily verified as an exercise that both sides of the first
$\iota$-reduction for $\nat$ have type $Q~\NO$ while they have type
$Q(\NS~t)$ for the second.}


\vspace*{-0.5cm}
\subsection{Parametric sorted algebras}
\label{ss:paralg}

{\bf Parametric sorted signature.}  Order-sorted algebras were
introduced as a formal framework for the OBJ language in
\cite{futatsugi85}, before to be generalized as {\em membership
equational logic} in \cite{bouhoula00tcs}. We use here a polymorphic
version of a restriction of the latter, by assuming given a signature
$(\Lambda,\Sigma)$, $\Lambda$ for the sort constructors, and $\Sigma$
for the function symbols made of a set of constructors for each sort
constructor, and of a set of defined symbols.  We shall use the
notation $f: \forall \ov{\alpha} .\, \sigma_1 \times \cdots \times
\sigma_n \ra \tau$ for symbol declarations.  As an example, we
describe natural numbers and parametric (non-dependent) list using an
OBJ-like syntax. We rule out here partiality, as introduced in
practice by destructor symbols, for sake of clarity.

We shall use $\cV=\{\alpha, \beta, \ldots\}$ for the set of sort
variables, and $\cT(\Sigma,\cV)=\{\sigma, \tau,
\ldots\}$ for the set of sort expressions.

{\small\begin{center}
\begin{tabular}{l@{\hspace*{10mm}}r}
\begin{tabular}{llcr}
\verb+sort+ & $\nat$ & : & $*$\\
\verb+sort+ & $\lst$ & : &  $* \ra *$\\
\verb+svar+ & $\alpha$ & : & $*$\\
\verb+cons+ & $\NO$ & : & $\nat$\\
\end{tabular}
&\begin{tabular}{llcr}
\verb+cons+ & $\NS$ & : & $\nat$ $\ra \nat$\\
\verb+fun+ & $\NP$ & : & $\nat \times \nat$ $\ra \nat$\\
\verb+cons+ & $\nil$ & : &  $\lst(\alpha)$\\
\verb+cons+ & $\cons$ & : & $\alpha \times \lst(\alpha)$ $\ra \lst(\alpha)$\\
\verb+fun+ & $@$ & : & $\lst(\alpha) \times \lst(\alpha)$ $\ra \lst(\alpha)$\\
\end{tabular}
\end{tabular}\end{center}}

\begin{definition}[Terms]
For any sort $\sigma $, let $\cX^\sigma$ be a countably infinite set
of {\em variables of sort $\sigma$}, s.t. all the $\cX^\sigma$'s are
pairwise disjoint. Let $\cX = \bigcup_{\sigma} \cX^\sigma$. For any $x
\in \cX$, we say that $x$ has sort $\sigma$ if $x \in \cX^\sigma$.
For any sort $\sigma$, the set $\cT_\sigma(\Sigma, \cX)$ of {\em terms
of sorts $\sigma$ with variables $\cX$} is the smallest set s.t.:
  \begin{enumerate}
  \item if $x \in \cX^\tau$, then $x \in \cT_\tau(\Sigma)$,
  \item if $t_1, \cdots, t_n \in \cT_{\sigma_1\xi}(\Sigma, \cX) \times
      \cdots \times \cT_{\sigma_2\xi}(\Sigma, \cX)$ where 
$f : \forall \ov{\alpha} .\, \sigma_1 \times \cdots
      \times \sigma_n \ra \tau$ and $\xi$ is a
      sort substitution,
    then $f(t_1,\ldots,t_n) \in \cT_{\tau\xi}(\Sigma, \cX)$.
  \end{enumerate}
Let $\cT(\Sigma, \cX)=\bigcup_{\sigma}(\cT_\sigma(\Sigma, \cX))$. A
term $t$ has sort $\sigma$ if $t \in \cT_\sigma(\Sigma, \cX)$.
\end{definition}

Note that the sets $\cX^\sigma$ play the role of a typing context.

\begin{example}
Assuming that $x$ is a variable of sort $\nat$, then $0$ and $0+x$ are
of sort $\nat$, while $\nil$ is of sort $\lst(\alpha)$, $\lst(\nat)$,
$\lst(\lst(\nat))$, etc.
\end{example}

\begin{definition}[Equations]
Equations $t=^\sigma u$ are pairs of terms of the same sort $\sigma$.
\end{definition}

\begin{example}
Assuming $x$ of sort $\nat$ and $l$ of sort $\lst(\lst((\nat))$,
$x+0=^{\nat}x$ is an equation of sort $\nat$ and
$cons(x,nil)=^{list(\nat)} car(l)$ is an equation of sort
$\lst(\nat)$.
\end{example}

We can therefore as usual build parametrized algebras for $\lst$,
algebras for $\nat$ and therefore get algebras for $\nat$,
$\lst(\nat)$, etc. Satisfaction of an equation in these algebras is
defined as usual.
In practice, type superscripts may be omitted when they can be infered
from the context.


\vspace*{-0.5cm}
\subsection{Embedding parametric algebras in \CIC}
\label{ss:embedding}

Our purpose here is to embed parametric multi-sorted algebra into
\CIC. As a result, two different, but related kinds of symbols will
coexist, in CIC and in the embedded algebraic sub-world. We shall
distinguish them by underlying symbols in CIC.

The first step of the translation maps, respectively sort constructors
and constructor symbols to \CIC inductive types and constructors.  We
start with natural numbers and its sort constructor
$\nat$. Constructor symbols of $\nat$ are simply all the constructors
symbols whose codomain is $\nat$, i.e. here $\NO$ and $\NS$. We thus
define $\uv{\nat}$ (the \CIC inductive type attached to $\nat$) as an
inductive type with two constructor types (one for $\NO$, and one for
$\NS$): $\uv{\nat} := \Ind(X : \Prop)\{ C_1(X), C_2(X) \}$.

The constructor types of $\uv{\nat}$ are simply the arities of $\NO$
and $\NS$ where $\nat$ is replaced with the constructor type variable:
$C_1(X) = X$ and $C_2(X) = X \ra X$. As expected, we obtain here the
standard inductive definition of natural numbers given in Section
\ref{ss:cic}: $\Ind(X : \Prop)\{ X, X \ra X \}$. The translation
$\uv{\NO}$ of $\NO$ (resp. $\uv{\NS}$ of $\NS$) is then simply
$\uv{\nat}^{[1]}$ (resp. $\uv{\nat}^{[2]}$).

Translating $\lst$ is not very different. Being of arity 1, with two
associated constructor symbols ($\nil$ and $\cons$), $\lst$ is mapped
to the already seen parametrized inductive type $\uv{\lst} = \AbsT{A :
T}{\Ind(\Prop)\{ X, A \ra X \ra X \}}$. Translation of constructors is
done the same way. We just need to care about curryfication of
symbols, and to replace sort variables with \CIC type
variables.\comment{For instance, the constructor type for $\cons :
\forall
\alpha .\, \alpha \times \lst(\alpha) \ra \lst(\alpha)$ is $\uv{\cons}
= \AbsT{A : *}{(\uv{\lst} \, A)^{[2]}}$.}

Finally, defined symbols are mapped to \CIC defined symbols, after
translating their type appropriately.


\vspace*{-0.5cm}
\subsection{Building in a first-order theory}
\label{ss:calculus}

We now start describing our new calculus \CCIC.\smallskip

{\bf Terms.} \CCIC uses the same set of sorts $\cS = \{ \Prop, \Type ,
\Extern\}$ and sets of variables $\cX = \cX^\Prop \cup \cX^\Type$ of
\CIC.  For any sort $\sigma \in \Lambda$, let $\cX_\sigma
\subseteq\cX^\Prop$ a infinite set of variables of sort $\sigma$ s.t.
$\{\cX_\sigma\}_\sigma$ is a family of pairwise disjoint sets. We also
assume that $\cX - \bigcup_\sigma \cX_\sigma$ is infinite.

Let $\cA = \{ \Ar, \Au \}$ a set of two constants, called {\em
annotations}, totally ordered by $\Au \prec_\cA \Ar$, where $\Ar$
stands for {\em restricted} and $\Au$ for {\em unrestricted}.  We use
$a$ for an arbitrary annotation. The role of annotations will be
explained later.

\begin{definition}[Pseudo-terms of $\CCIC$]
Given a parametric sorted signature $(\Lambda,\Sigma)$, the algebra
$\cL$ of {\em pseudo-terms} of \CCIC is defined as:

  \noindent\centering
  \begin{tabular}{r@{$\;$}c@{$\;$}l}
    $t, u, T, U, \ldots$ & $:=$
      & $s \in \cS \vertrel
         x \in \cX \vertrel
         \ProdT {x :^a T}{t} \vertrel
         \AbsT  {x :^a T}{t}
      \vertrel t \, u \vertrel
         f \in \Sigma \vertrel \sigma \in \Lambda$ \\[.3em]
        & $\vertrel$ & $\ccieq \vertrel \Eq_T(t)
      \vertrel\Ind(X : t)\{ \vv{T_i} \}
      \vertrel t^{[n]} \vertrel
         \ElimT{t}{T}{\vv{u_i}}{U}{\vv{w_j}}$
  \end{tabular}
\end{definition}

In order to make definitions more convenient, we assume in the
following that $\Lambda$ contains the symbols $\ccieq, \nat$ and
$\lst$, and that $\Sigma$ contains the symbols $\NO, \NS$ and $\Eq$.

Compared with \CIC, the differences are:
\begin{itemize}
\item the internalization of the first-order symbols,
\item the internalization of the equality predicate:\\
- $t \ccieq_T u$ denotes the
 equality of the two terms (of type $T$) $t$ and $u$,\\
- $\Eq_T(t)$ represents the reflexivity proof of $t
    \ccieq_T t$.
\item annotations in products and abstractions are used
to control the formation of applications as it can be seen from the
new {\rulename{App}} rule given at Figure \ref{fig:inf-ccic-rules}.
\end{itemize}

\begin{notation}
When $x$ is not free in $t$, $\ProdT{x :^a T}{t}$ is written $T
\ra^a t$. The default annotation, when not specified in a product or
abstraction, is the {\em unrestricted} one.
\end{notation}

As usual, there is a layered set of syntactic classes for $\cL$:

\begin{definition}[Syntactic classes]
The pairwise disjoint syntactic classes of \CCIC called {\em objects}
($\cO$), {\em predicates} ($\cP$), {\em kinds} ($\cK$), {\em kinds
  predicates} ($\cM$), and $\Extern$ are defined as usual:\smallskip

{\small
\noindent
$-~\cO ::= \cX^\Prop     \vertrel
              f \in \Sigma \vertrel
              \cO\,\cO     \vertrel
              \cO\,\cP     \vertrel
              \AbsT{x^\Prop :^a \cP}{\cO} \vertrel
          \AbsT{x^\Type :^a \cK}{\cO}
          \vertrel \ElimT{\cO}{\cP}{\vv{\cO}}{\cO}{\vv{\cO}}$\\
$-~\cP ::= \cX^\Type           \vertrel
              \sigma \in \Lambda \vertrel
              \cP\,\cO           \vertrel
              \cP\,\cP           \vertrel
              \AbsT{x^\Prop :^a \cP}{\cP} \vertrel
             \AbsT{x^\Type :^a \cK}{\cP}$\\
           \hspace*{1cm}$ \vertrel
              \ElimT{\cO}{\cP}{\vv{\cO}}{\cP}{\vv{\cP}}
\vertrel  \ProdT{x^\Prop :^a \cP}{\cP} \vertrel
              \ProdT{x^\Type :^a \cK}{\cP}$\\
$-~\cK ::= \Prop      \vertrel
              \cK \, \cO \vertrel
              \cK \, \cP \vertrel
              \AbsT{x^\Prop :^a \cP}{\cK} \vertrel
              \AbsT{x^\Type :^a \cK}{\cK} \vertrel
            \ProdT{x^\Prop :^a \cP}{\cK} \vertrel
              \ProdT{x^\Type :^a \cK}{\cK}$\\
$-~\cM ::= \Type      \vertrel
            \ProdT{x^\Prop :^a \cP}{\cM} \vertrel
              \ProdT{x^\Type :^a \cK}{\cM}$\\
$-~\Extern ::= \Extern$\\}
\smallskip
\noindent
This enumeration defines a successor function $+1$ on classes ({\small
  $\cO+1 = \cP$, $\cP+1 = \cK$, $\cK+1=\cM$, $\cM+1=\Extern$}). We
also define $\Class(t) = \cD$ if $t\in \cD$ and $\cD \in \{ \cO, \cP,
\cK, \cM, \Extern \}$.
\end{definition}

From now on, we only consider {\em well-constructed terms} (i.e.
terms whose class is not $\bottom$) and {\em well-constructed
  substitution} (i.e. substitutions s.t. $\Class(x) = \Class(x\theta)$
for any $x$ in its domain).  It is easy to check that if $t$ is a
well-constructed term and $\theta$ a well-constructed substitution,
then $\Class(t) =\Class(t\theta)$. It is also well-known that
$\rw{\beta\iota}$-reduction preserves term classes.


\begin{definition}[Pseudo-contexts of \CCIC]
The typing environments of \CIC are defined as $\Gamma, \Delta ::= []
\vertrel \Gamma, [x :^a T]$ s.t. a variable cannot be declared twice.
We use $\dom(\Gamma)$ for the domain of $\Gamma$ and $x\Gamma$ for the
type associated to $x$ in $\Gamma$.
\end{definition}

The rules defining the \CCIC typing judgment $\Gamma \vdash t : T$
are the same as for \CIC except the rules for application and
conversion given at Figure \ref{fig:inf-ccic-rules}.

\begin{figure}
\begin{tabular}{cc}
\begin{minipage}{6cm}\centering
\InferRule{App}
          {\Gamma \vdash t : \ProdT{x :^a U}{V} \quad
           \Gamma \vdash u : U \\
           \Trm{if $a = \Ar$ and $U \rwrefltr{\beta} t_1 \ccieq_T t_2$
                with $t_1, t_2 \in \cO$}\\
           \Trm{then $t_1 \cnv{\Gamma} t_2$ must hold}}
          {\Gamma \vdash t\,u : V \{ x \mapsto u \}}
\end{minipage}
&\begin{minipage}{6cm}\centering
\InferRule{Conv}
  {\Gamma \vdash t : T \quad \Gamma \vdash T' : s' \quad T \cnv{\Gamma} T'}
  {\Gamma \vdash t : T'}
 \end{minipage}
\end{tabular}
\smallskip
\caption{\label{fig:inf-ccic-rules} \CCIC modified typing rules}
\end{figure}

\vspace*{-1.0cm}
\subsection{Conversion}
\label{ss:conversion}

We are now left with defining the conversion relation $\cnv{\Gamma}$,
whose definition needs some preparation, since:
\begin{itemize}
\item conversion is defined on CCIC terms, but the first-order
decision procedures operate on algebraic terms. We therefore need to
translate CCIC terms into algebraic terms, a process we call
{\em algebraisation}.
\item conversion will operate on weak terms only, a notion introduced
in Section \ref{s:weak-terms}. Non-weak terms will be converted with
$\beta\iota$-reduction only, to forbid lifting up inconsistencies from
the object level to the type level. This is crucial to avoid breaking
strong normalization, and therefore decidability of type-checking in
presence of inconsistent user's assumptions.
\end{itemize}

{\bf Algebraisation.} Our calculus has a complex notion of computation
reflecting its rich structure made of three ingredients: the typed
lambda calculus, the inductive types with their recursors and the
integration of the first order theory $\cT$ in its conversion.  To
achieve this integration, goals are sent to the first order theory
$\cT$ together with a set of proof hypotheses extracted from the
current context.

Algebraisation is the first step of this extraction: it allows
transforming a \CCIC term into its first-order counterpart.  We
illustrate this with an example, $\cT$ being Presburger's arithmetic.

We begin by the simplest case, directly taken from \CCNAT, the
extraction of pure algebraic, non parametric, equations. Suppose that
the proof environment contains equations of the form $c \ccieq 1 + d$
and $d \ccieq 2$ with $c$ and $d$ variables of sort $\nat$. What is
expected is that the set of hypotheses sent to the theory $\cT$
contains the two well formed $\cT$-formulas $c = 1+d$ and $d =
2$. This leads to a first definition of equations extraction:
\begin{enumerate}
\item a term is algebraic if it is of the form $0$, or $S \, t$, or $t
  + u$, or $x \in \cX_\sN$. The {\em algebraisation} $\cA(t)$ of an
  algebraic term is then defined by induction: $\cA(0) = 0$,
  $\cA(S \, t) = S(\cA(t))$, $\cA(t + u) = \cA(t) + \cA(u)$ and
  $\Alg(x_\sN) = x_\sN$,
\item a term is an extractable equation if it is of the form $t \ccieq
  u$ with $t$ and $u$ algebraic terms. The extracted equation is then
  $\cA(t) = \cA(u)$.
\end{enumerate}

The definition becomes harder for parametric signatures. The theory of
lists gives us a paradigmatic example.  From the definition of
embedding a polymorphic multi-sorted algebra into \CIC, we know that
the symbol $@$ has {\small$\ProdT{T : \Prop}{\lst\,T \ra \lst\,T \ra
\lst\,T}$} for type. Thus, a fully applied, well formed term having
the symbol $@$ at head position must be of the form $(@ \, T \, l1 \,
l2)$, $T$ being the type of the elements of the lists $l1$ and
$l2$. Algebraisation of such a term will erase all type parameters: in
our example, $\Alg(@\, T \, l1 \, l2) = @(\Alg(l1),\Alg(l2))$.

Algebraisation of non-pure algebraic terms is done by abstracting
non-algebraic subterms with fresh variables. For example,
algebraisation of $1 + t$ with $t$ non-algebraic will lead to $1 +
x_\nat$ where $x_\nat$ is an abstraction variable of sort $\nat$ for
$t$. Of course, if the proof context contains two equations of the
form $c \ccieq 1 + t$ and $d \ccieq 1 + u$ with $t$ and $u$
$\beta\iota$-convertible, $t$ and $u$ should be abstracted by a unique
variable so that $c = d$ can be deduced in $\cT$ from $c = 1 + y_\nat$
and $d = 1 + y_\nat$. The problem is harder for:
\begin{itemize}
\item {\em parametric symbols}: in $(\cons \, T \, t \, (\nil \, U))$
  with $t$ non algebraic, should $t$ be abstracted by a variable of
  sort $\nat$ or $\lst(\nat)$ ?
\item {\em ill-formed terms}: should $(\cons \, T \, 0 \, (\cons \, T
  \, (\nil \, U) \, (\nil \, T)))$ be abstracted as a list of natural
  numbers or as a list of lists ?
\end{itemize}

\noindent Our solution is to postpone decisions:
$\Alg(t)$ will be a function from $\Lambda$ to the terms of $\cT$
s.t. $\Alg(t)(\sigma)$ is the algebraisation of $t$ under the
condition that $t$ is a \CCIC representation of a first order term of
sort $\sigma$.

We now give the formal definition of $\Alg(\cdot)$. We assume:

- a $\Lambda$-sorted family $\{ \cY_\sigma \}_\sigma$ of
pairwise disjoint countable infinite sets of variables of sort
$\sigma$. Let $\cY = \bigcup_\sigma \cY_\sigma$;

- for any equivalence relation $\cR$ and sort $\sigma \in \Lambda$,
we assume a function $\pi_\cR^\sigma : \CCIC(\cX)
\ra \cY_\sigma$ s.t. $\pi_\cR^\sigma(t) = \pi_\cR^\sigma(u)$ if and
only if $t \mathrel{\cR} u$ (i.e. $\pi_\cR^\sigma(t)$ is the element
of $\cY_\sigma$ representing the class of $t$ modulo $\cR$).

\begin{definition}[Well applied term]
A term is well applied if it is of the form $f \,
[\vv{T_\alpha}]_{\alpha \in \ov{\alpha}} \, t_1 \,\cdots\, t_n$ with
$f : \forall \ov{\alpha} .\, \sigma_1 \times \cdots \times \sigma_n
\ra \sigma$.
\end{definition}

\begin{example}
Example of well applied terms are $0$, $S \, t$, or $\cons \, T \, x
\, l$, ~$T$ being the type parameter here.  Note that we do not
require the term to be well formed.
\end{example}

In case of partial symbols, such as $\car$ for lists, this definition
must be changed slightly by adding a new argument, the proof that the
input satisfies the appropriate guard, here that it is not $\nil$.

\begin{definition}[Algebraisation]
The {\em algebraisation of $t\in \CCIC$ modulo an equivalence relation
$\cR$} is the function $\Alg_\cR(t) :
\Lambda \ra \cT(\cX^\Prop \cup \cY)$ defined by:

  $
  \begin{array}{rcl}
    \Alg_\cR(x_\sigma)(\sigma) & = &
         x_\sigma\\
    \Alg_\cR(f \, \vv{T} \, [u_i]_{i \in n})(\tau\xi) & = &
        f(\Alg_\cR(u_1)(\sigma_1\xi),\ldots,
        \Alg_\cR(u_n)(\sigma_n\xi))\\
    \Alg_\cR(t)(\tau) & =&
         \pi_\cR^\tau(t) \quad \text{otherwise}\\
  \end{array}
  $

\noindent
where $f:\forall \ov{\alpha} .\, \sigma_1
            \times \cdots \sigma_n \ra \sigma$, 
$f \, \vv{T} \, [u_i]_{i \in n}$~ is
            well applied, and 
$\xi$ is a $\Lambda$-substitution.

For any relation $R$, $\Alg_R$ is defined as $\Alg_\cR$ where $\cR$ is
the smallest equivalence relation containing $R$.  We call {\em
$\sigma$-alien} (or {\em alien} when the context is clear) a subterm
of $t$ abstracted by a variable in $\cY_\sigma$, and say that $t$ is
{\em algebraic} w.r.t. $\sigma$ if contains no $\sigma$-alien. We
denote by $\Algt_\sigma$ the set of algebraic terms w.r.t. $\sigma$,
and by $\Algt=\bigcup_{\sigma\in\Lambda}\Algt_\sigma$ the set of
algebraic terms.
\end{definition}

\begin{example}
Let $t \equiv \cons \, T \, \NO \, (\cons \, U \, (\nil \, V) \, (\nil
\, U))$, $R$ be a relation on $\CCIC$ terms, $\sigma
= \lst(\nat)$, and $x_\nat, y_\lst, z_\nat, x_\alpha$ and $y_\alpha$
be abstraction variables. Then:
{\small
  \begin{align*}
    \Alg_R(t)(\sigma)
    & = \cons(
     \Alg_R(\NO)(\nat),
     \Alg_R(\cons \, U (\nil \, V) \, (\nil \, U))(\sigma))\\
    & = \cons(0, \cons(\Alg_R(\nil \, V)(\nat),
                       \Alg_R(\nil \, U)(\sigma)))
     = \cons(0, \cons(x_\nat, \nil))\\
    \Alg_R(t)(\lst(\sigma))
    & = \cons(
      \Alg_R(\NO)(\sigma), \Alg_R(\cons
       U (\nil \, V) \, (\nil \, U))(\lst(\sigma)))\\
    & = \cons(
      y_\lst, \cons(\Alg_R(\nil \, V)(\sigma),
     \Alg_R(\nil \, U)(\lst(\sigma))))
    = \cons(y_\lst, \cons(\nil, \nil))
  \end{align*}
}
\noindent
$\Alg_R(t)(\lst(\alpha)) = \cons(x_\alpha, \cons(y_\alpha, \nil))$
and $\Alg_R(t)(\nat) = z_\nat$.

It is clear from the above example that the algebraisation of a term
depends on the expected sort of the result: when abstracting the
(heterogeneous and ill-formed) list $0 :: \nil :: \nil$ as a list of
lists, $0$ is seen as an alien which must be abstracted.  When this
list is abstracted as a list of natural numbers or as a polymorphic
list, $0$ is considered algebraic and the first occurrence of $\nil$
as an alien to be abstracted.  Finally, if the list is algebraised as
a natural number, it is abstracted by a variable.
\end{example}


{\bf Weak terms.}
\label{s:weak-terms}
We first distinguish a class of terms called {\em weak}. This class of
terms will play an important role in the following as they restrict
the interaction between the conversion at object level and the strong
$\iota$-reduction.

An example of non weak term is
\begin{center}
$t = \AbsT{x : \nat}{\ElimTS{x}{\nat}{}{Q}
{\nat,\AbsTT{x : \nat}{T : Q\,x}{\nat \ra \nat}}}$
\end{center}
Such a term is problematic in the sense that when applied to
convertible terms, it can $\beta\iota$-reduce to type-level terms that
are not $\beta\iota$-convertible.  Suppose that the conversion
relation is canonically extended to $\CCIC$. Assume a typing
environment $\Gamma$ s.t. $\NO \cnv{\Gamma} \NS\,\NO$, and hence, by
congruence, $t\,\NO \cnv{\Gamma} t\,(\NS\,\NO)$. Now, it is easy to
check that $t\,\NO \rwrefltr{\beta\iota} \nat$ and $t\,(\NS\,\NO)
\rwrefltr{\beta\iota} (\nat \ra \nat)$.  Strong normalization of
$\beta$-reduction is then broken by encoding the term $\omega =
\AbsT{x : \nat}{x\,x}$.

In contrast, {\em weak} terms lift no inconsistencies from
object level to a higher level:

\begin{definition}[Weak terms]
  A term is {\em weak} if it contains no
  \begin{paraenum}
  \item  applied type-level variable, and
  \item 
term of the form $\ElimT{t}{I}{\vv{u}}{Q}{\vv{f}}$ with $t$ open.
  \end{paraenum}
\end{definition}

{\bf Extractable terms.}
\label{s:extractable-terms}
From now on, let $\cO^+$ be an arbitrary set of \CCIC terms. This set
will be used in the conversion definition to restrict the set of
{\em extractable equations} of a given environment: only equation of
the form $t \ccieq u$ with $t$ and $u$ in $\cO^+$ will be considered.

At the moment, we only require $\cO^+$ to be a subset of $\cO$. Note
that taking $\cO^+ = \cO$ does not compromise the standard calculus
properties (subject reduction, type unicity, strong normalization of
$\beta\iota$-reduction, $\ldots$) but the decidability. E.g., if $\cT$
is the Presburger arithmetic, allowing the extraction of
\begin{center}
$\AbsT{x :^a \nat}{f\,x} \ccieq \AbsT{x :^a \nat}{f\,(x \NP 2)}$
\end{center}
would require -~for checking conversion~- to decide any
statement of the form
\begin{center}
  $\cT \vDash (\forall x .\, f(x) = f(x+2)) \ra t = u$,
\end{center}
which is well known to be impossible.\smallskip


{\bf Conversion relation.} We have now all necessary ingredients to
define our conversion relation $\cnv{\Gamma}$:

\begin{definition}[Conversion relation]
Rules of Figure \ref{fig:ccic-conversion-relation} define a family $\{
\cnv{\Gamma} \}$ of \CCIC binary relations indexed by a
(non-necessarily well-formed) context $\Gamma$.
\end{definition}

\begin{figure}
\centering
\begin{tabular}{cc}
\begin{minipage}{4cm}\centering
  \InferRule{Refl}{ }{t \cnv{\Gamma} t}
\end{minipage}
&\begin{minipage}{6cm}\centering
  \InferRule{Eq}
    {[x :^\Ar T] \in \Gamma \quad
     T \rwrefltr{\beta\iota} t \ccieq u \quad
     t, u \in \cO^+}
    {t \cnv{\Gamma} u}
 \end{minipage}
\end{tabular}\\[2mm]

\begin{tabular}{cc}
\begin{minipage}{5cm}\centering
  \InferRule{Lam}
    {T \cnv{\Gamma} U \quad t \cnv{\Gamma, [x :^a T]} u}
    {\AbsT{x :^a T}{t} \cnv{\Gamma} \AbsT{x :^a U}{u}}
\end{minipage}
&\begin{minipage}{5cm}\centering
  \InferRule{Prod}
    {T \cnv{\Gamma} U \quad t \cnv{\Gamma, [x :^a T]} u}
    {\ProdT{x :^a T}{t} \cnv{\Gamma} \ProdT{x :^a U}{u}}
\end{minipage}
\end{tabular}\\[2mm]

\begin{tabular}{cc}
\begin{minipage}{4cm}\centering
  \InferRule{$\beta\iota$-Left}
    {t \rw{\beta\iota} t' \quad t' \cnv{\Gamma} u}
    {t \cnv{\Gamma} u}
\end{minipage}
&\begin{minipage}{6cm}\centering
  \NInferRule
    {\mbox{$t, t', f, f'$ are weak}\\
     t \rwequiv{\beta\iota} t' \quad I \cnv{\Gamma} I' \quad Q \cnv{\Gamma} Q' \quad
     \vv{v} \cnv{\Gamma} \vv{v}' \quad \vv{f} \cnv{\Gamma} \vv{f}'}
    {\ElimT{t}{I}{\vv{v}}{Q}{\vv{f}} \cnv{\Gamma} \ElimT{t'}{I'}{\vv{v}'}{Q'}{\vv{f}'}}
 \end{minipage}
\end{tabular}\\[2mm]

\begin{tabular}{cc}
\begin{minipage}{4cm}\centering
  \InferRule{$\beta\iota$-Right}
    {u \rw{\beta\iota} u' \quad t \cnv{\Gamma} u'}
    {t \cnv{\Gamma} u}
\end{minipage}
&\begin{minipage}{6cm}\centering
  \InferRule{App$^\cW$}
    {t_1 \cnv{\Gamma} u_1 \quad t_2 \cnv{\Gamma} u_2 \quad \text{$t_i, u_i$ are weak}}
    {t_1\,t_2 \cnv{\Gamma} u_1\,u_2}
 \end{minipage}
\end{tabular}\\[2mm]

\begin{tabular}{cc}
\begin{minipage}{6cm}\centering
  \InferRule{Ded}
    {E \vDash \Alg_{\cnv{\Gamma}}(t)(\tau) = \Alg_{\cnv{\Gamma}}(u)(\tau) \quad t, u \in \cO^+\\
     \begin{array}{c} E = \{\Alg_{\cnv{\Gamma}}(w_1)(\sigma) = \Alg_{\cnv{\Gamma}}(w_2)(\sigma) \\\hspace*{1cm}\vertrel
w_1 \cnv{\Gamma} w_2, \sigma \in \Lambda, w_1, w_2 \in \cO^+\} \end{array}
    }
    {t \cnv{\Gamma} u  \quad [\Gamma, t, u]}
 \end{minipage}
\end{tabular}
  \smallskip
  \caption{\label{fig:ccic-conversion-relation} \CCIC conversion relation}
\end{figure}

Note that the rule {\rulename{DED}} performing deductions in the first
order theory, here Presburger arithmetic, outputs a certificate
$[\_,\_,\_]$ made of the environment and the two terms to be proved
equivalent under this environment, each time it is called. While this
certificate must depend on these three data, it may of course carry
additional information depending on the considered first-order theory.

The main differences with the calculus \CCNAT defined
in \cite{blanqui07csl} are the following:
  \begin{itemize}
  \item The \rulename{App} rule has been split into two rules:
    \rulename{App$^\cS$} and \rulename{App$^\cW$}. Conversion for
    strong terms is restricted to $\beta\iota$-conversion.

  \item Conversion for the first argument of an
 $\Elim$ is restricted to $\beta\iota$-conversion.

  \item The rules for transitivity and symmetry have been removed,
    which eases the proofs, notably that the deduction part of the
    conversion relation works at object level only. We prove later
    that the conversion relation is transitive and symmetric on well
    formed terms, thus recovering type unicity.

  \item The rules for $\beta\iota$-conversion perform one
    reduction step only, which also eases proofs. Therefore
    $u\rwequiv{\beta\iota}v$ should be understood as $\exists w$
    s.t. $u\rw{\beta\iota} w$ and $v\rw{\beta\iota}w$.
  \end{itemize}


\vspace*{-0.5cm}
\subsection{Decidability of type-checking}

\CCIC enjoys all needed meta-theoretical properties (strong
normalization, confluence, subject reduction), and therefore
consistency follows:

\begin{theorem}
There is no proof of $\ProdT{x : \Prop}{x}$ in the empty environment.
\end{theorem}

All proofs are similar to those made for PTSs with the same succession
of meta-theoretical lemmas, but need more preparation. This is in
particular the case with the substitution lemma which is much harder
than usual.

As said, type-checking in a dependent type theory is non-trivial,
since the rule $\rulename{Conv}$ is not syntax-oriented. The classical
solution to this problem is to eliminate $\rulename{Conv}$ and replace
$\rulename{App}$ by the following rule.
The proof is not difficult.

\[\InferRule{App}
          {\Gamma \vdash t : \ProdT{x :^a U}{V} \quad
           \Gamma \vdash u : U' \quad U \cnv{\Gamma} U' \\
           \Trm{if $a = \Ar$ and $U \rwrefltr{\beta} t_1 \ccieq_T t_2$
                with $t_1, t_2 \in \cO$}
           \Trm{then $t_1 \cnv{\Gamma} t_2$ must hold}}
          {\Gamma \vdash t\,u : V \{ x \mapsto u \}}\]

Decidability of type-checking in $CCIC$ therefore reduces to
decidability of $\cnv{\Gamma}$, the environment $\Gamma$ being
arbitrary, possibly containing ill-formed terms or even being
inconsistent. To show that $\cnv{\Gamma}$ is decidable, we proceed as
previously, by modifying the definition in order to make it
syntax-oriented: we show that two arbitrary terms are convertible iff
their $\beta\iota$-normal forms are convertible by the syntax-oriented
{\em weak convertibility} relation $\cnvbl{\Gamma}$ given at Figure
\ref{fig:soc-ccic}, in which, to any environment $\Gamma$, we
associate the set $\Eq(\Gamma) = \{ t = u \vertrel [x :^\Au T] \in
\Gamma, x\Gamma \rwrefltr{} t \ccieq u, t, u \in \cA \}$.

\begin{lemma}
Given $\Gamma$ an environment and $t,u$ two terms,
$t\cnv{\Gamma} u$ iff
$\nf{t}{\beta\iota}\cnvbl{\Gamma}
\nf{u}{\beta\iota}$.
\end{lemma}

This is the main technical result of the decidability proof, which
proceeds by induction on the definition of $\cnv{\Gamma}$.  Note that
the numerous conditions of the form $\cT, \Eq(\Gamma) \not\vDash 0 =
1$ in the rules defining $\cnvbl{\Gamma}$ are required to make them
mutually exclusive.

\begin{figure}
\centering
\begin{tabular}{c@{\quad}c@{\quad}c}
  \InferRule{Refl-$\Prop$}{ }{\Prop \cnvbl{\Gamma} \Prop}
&  \InferRule{Refl-$\Type$}{ }{\Type \cnvbl{\Gamma} \Type}

&  \InferRule{Refl-$\cX$}
   {x \in \cX \quad
    \mbox{$\cT, \Eq(\Gamma) \not\vDash 0 = 1$ or $x \not\in \cX^\Prop$}}
   {x \cnvbl{\Gamma} x}
\end{tabular}\\[2mm]

\begin{tabular}{cc}
  \InferRule{Unsat}
    {t, u \in \cO \quad \cT, \Eq(\Gamma) \vDash 0 = 1}
    {t \cnvbl{\Gamma} u}

&  \quad\InferRule{Lam}
    {T \cnvbl{\Gamma} U \quad t \cnvbl{\Gamma,[x :^a T]} u \\
     \cT, \Eq(\Gamma) \not\vDash 0 = 1 \mbox{ or }\\
     \mbox{$\AbsT{x :^a T}{t}$ and $\AbsT{x :^a U}{u}$ not in $\cO$}}
    {\AbsT{x :^a T}{t} \cnvbl{\Gamma} \AbsT{x :^a U}{u}}
\end{tabular}\\[2mm]

\begin{tabular}{cc}
  \InferRule{Prod}
    {T \cnvbl{\Gamma} U \quad t \cnvbl{\Gamma, [x :^a T]} u}
    {\ProdT{x :^a T}{t} \cnvbl{\Gamma} \ProdT{x :^a U}{u}}

&  \InferRule{$\cW$}
    {t = t' \quad I \cnvbl{\Gamma} I' \quad Q \cnvbl{\Gamma} Q' \quad
     \vv{v} \cnvbl{\Gamma} \vv{v}' \quad \vv{f} \cnvbl{\Gamma} \vv{f}' \\
     \mbox{$t, t', \vv{f}, \vv{f}'$ are weak}
     \cT, \Eq(\Gamma) \not\vDash 0 = 1 \mbox{ or }\\
\mbox{$\Elim(t,\ldots)\{\ldots\}$ and $\Elim(t',\ldots)\{\cdots\}$ not in $\cO$} \\
}
    {\ElimT{t}{I}{\vv{v}}{Q}{\vv{f}} \cnvbl{\Gamma} \ElimT{t'}{I'}{\vv{v}'}{Q'}{\vv{f}'}}
\end{tabular}\\[2mm]

\begin{tabular}{cc}
\begin{minipage}{5cm}
  \InferRule{App$^\cS$}
    {t_1 \equiv u_1 \quad t_2 \equiv u_2 \\
     \mbox{$\cT, \Eq(\Gamma) \not\vDash 0 = 1$ or}\\
     \mbox{$t_1\,t_2$ and $u_1\,u_2$ not in $\cO$}\\
     \mbox{$t_1\,t_2$ or/and $u_1\,u_2$ is not weak}}
    {t_1\,t_2 \cnvbl{\Gamma} u_1\,u_2}\\[2mm]

  \InferRule{App$^\cW$}
    {t_1 \cnvbl{\Gamma} u_1 \quad t_2 \cnvbl{\Gamma} u_2 \quad \text{$t_i, u_i$ weak} \\
     \mbox{$\cT, \Eq(\Gamma) \not\vDash 0 = 1$ or}\\
     \mbox{$t_1\,t_2$ and $u_1\,u_2$ not in $\cO$}}
    {t_1\,t_2 \cnvbl{\Gamma} u_1\,u_2}
\end{minipage}
&  \InferRule{Ded}
    {\cT, \Eq(\Gamma) \not\vDash 0 = 1) \\
    t = C_t[a_1,\ldots,a_k] \quad u = C_u[a_{k+1},\ldots,a_{k+l}] \\
     \mbox{$C_t$ or $C_u$ is a non-empty algebraic context} \\
     \mbox{all the $a_i$'s have empty algebraic caps} \\
     \mbox{the $c_i$'s are fresh constants s.t. $c_i = c_j$ iff $a_i \cnvbl{\Gamma} b_j$} \\
     \cT, \Eq(\Gamma) \vDash C_t[c_1,\ldots,c_k] = C_u[c_{k+1},\ldots,c_{k+l}]}
    {t \cnvbl{\Gamma} u}
\end{tabular}

\caption{\label{fig:soc-ccic} \CCIC syntax-oriented conversion}
\end{figure}

\begin{example}
Let $\Gamma = [c : \nat], [p :^\Ar (\AbsT{x : \nat}{x}) \, \NO \ccieq
c]$. Then $(\AbsT{x : \nat}{x + x})\,\NO \cnvbl{\Gamma} c$ and
$(\AbsT{x : \nat}{x + x})\,\NO \cnvbl{\Gamma} c$, using congruence and
deduction of $\cnv{\Gamma}$ and $\cnvbl{\Gamma}$.

In contrast, $\beta$-reducing $(\AbsT{x : \nat}{x + x})\,\NO$ yields
$\NO \NP \NO \cnv{\Gamma} c$, but not $\NO \NP \NO
\cnvbl{\Gamma} c$. Indeed, $(\AbsT{x : \nat}{x \NP x})\,\NO$ and $\NO
\NP \NO$ are no more $\cnvbl{\Gamma}$-convertible, a direct
consequence of removing $\beta\iota$-reduction from $\cnv{\Gamma}$:
the equation $(\AbsT{x : \nat}{x}) \, \NO \ccieq c$ cannot be used
anymore, since $\NO \NP \NO$ is not $\cnvbl{\Gamma}$ convertible to
$(\AbsT{x : \nat}{x}) \, \NO$).

Now, normalizing all terms as well as the environment $\Gamma$, we can
recover convertibility for $\cnvbl{}$: $\NO \NP \NO \cnvbl{\Gamma_{\da
\beta\iota}} c$, the extractable equation of $\Gamma_{\da \beta\iota}$
being now $\NO \ccieq c$.
\end{example}

As a consequence, we obtain:

\begin{theorem}
$\cnv{\Gamma}$ is decidable for any environment $\Gamma$ when taking
for $\cO^+$ the set of terms that are reducible to an algebraic terms.
\end{theorem}

\noindent
and therefore, our main result follows:

\begin{theorem}
The type-checking relationship $\Gamma \vdash t:T$ is decidable in
$CCIC$.
\end{theorem}


\vspace*{-0.5cm}
\section{Using \CCIC}
\label{s:usingccic}

We give here a detailed example illustrating the advantages of \CCIC,
based on the inductive type of words introduced in Section
{}\ref{ss:cic}.\smallskip

{\bf In Coq.} First, we give a development in Coq, therefore based on
\CIC.\comment{Indentations have been modified by hand to make it fit.}

{\scriptsize\begin{verbatim}
Variable T : Set.

Inductive word : nat -> Set :=
| epsilon : word 0
| char : T -> word 1
| append : forall n p, word n -> word p -> word (n+p).

Lemma plus_n_0_transparent : forall n, n+0=n.
Proof. induction n as [| n IHn]; simpl;
  [idtac | rewrite -> IHn]; trivial. Defined.

Lemma plus_n_Sm_transparent: forall n m, n+(S m)=S(n+m).
Proof. intros n m; induction n as [| n IHn];
  simpl; [idtac | rewrite -> IHn]; trivial. Defined.

Lemma plus_assoc_transparent: forall n p q, (n+p)+q=n+(p+q).
Proof. intros n p q; elim n; [trivial | intros k].
  simpl; intros H; rewrite -> H; trivial. Defined.

Definition reverse_acc : forall n, word n -> forall p, word p -> word (p+n).
Proof. intros n wn; induction wn as [| c | n p wn IHwn wp IHwp];
  intros k wk. rewrite plus_n_0_transparent; exact wk.
  rewrite plus_n_Sm_transparent; rewrite plus_n_0_transparent;
    exact (append (char c) wk).
  rewrite <- plus_assoc_transparent; exact (IHwp _ (IHwn _ wk)). Defined.

Fixpoint reverse n (w : word n) {struct w} : word n :=
match w in word k return word k with
| epsilon => epsilon
| char c => char c
| append n1 n2 w1 w2 => reverse_acc w2 w1 end.
\end{verbatim}}
\comment{
Fixpoint to_Tlist n (w : word n) {struct w} : list T
:= match w with
| epsilon => nil
| char c => c :: nil
| append n1 n2 w1 w2 => to_Tlist _ w1 ++ to_Tlist _ w2
end.

Variables a b c : T.
Eval compute in (to_Tlist (reverse
  (append (append epsilon (char a)) (append (char b) (char c))))).

(* Print c :: b :: a :: nil *)
\end{verbatim}}

The example of {\em palindromes} as words satisfying the property
{\small\verb#word_eq m reverse m#} is carried out in Strub's thesis
(see his website). It yields a much more complex Coq development than
the above, since it involves the equality over (quotients) of
words.\smallskip

{\bf In \CCIC.} We now make the similar development in \CCIC, using a
self-explanatory syntax. The definition of \verb#reverse# reduces then to:

\comment{
Variable T : Set.

Inductive word : nat -> Set :=
| epsilon : word 0
| char    : T -> word 1
| append  : forall n p, word n -> word p -> word (n+p).
}
{\scriptsize\begin{verbatim}
Fixpoint reverse n (w : word n) {struct w} : word n := match w with
| epsilon          => epsilon
| char c           => char c
| append _ _ w1 w2 => append (reverse w2) (reverse w1) end.
\end{verbatim}}

Typing of the third clause of reverse will use here Presburger's
arithmetic, since
{\small\verb#append n1 n2 w1 w2#}
has type
{\small\verb#word (n1 + n2)#},
while {\small\verb#append n2 n1 w2 w1#}
has type
{\small\verb#word (n2 + n1)#}, two types that are not
convertible in \CIC, but which become convertible in \CCIC.
We can easily see  with this example the immense benefit brought by
internalizing Presburger's arithmetic. Note that a
single certificate is generated for this conversion:\\
{\small\verb#[n1 : nat, n2: nat, w1 : word n1, w2: word n2, n1 + n2, n2 + n1]#}

\comment{We now introduce palindromes and prove that reverse is the identity on
palindromes.

{\scriptsize\begin{verbatim}
Variables M : Set.
Variables f : forall n, word n -> M.

Definition word_eq n1 n2 (w1 : word n1) (w2 : word n2) := (f w1) = (f w2).
Notation "w1 == w2" := (word_eq w1 w2) (at level 70, no associativity).

Lemma reverse_invol : forall n (w : word n), reverse (reverse w) = w.
Proof. intros w; induction w; simpl; trivial.
  rewrite -> IHw1; rewrite -> IHw2. trivial. Qed.

Definition is_pal n (w : word n) := w == (reverse w).

Lemma is_pal_reverse : forall n (w : word n), is_pal w -> is_pal (reverse w).
Proof. intros n w H. unfold is_pal; unfold is_pal in H.
  unfold word_eq; unfold word_eq in H. rewrite <- H;
    rewrite -> (reverse_invol w). trivial. Qed.
\end{verbatim}}
}

\vspace*{-0.5cm}
\section{Conclusion}
\label{c:conc}

{}\CCIC is an extension of CIC by arbitrary first-order decision
procedures for equality. We have shown here with a detailed example
using Presburger's arithmetic the benefit of the approach with respect
to the current implementation of Coq based on \CIC: more terms can be
typed especially in presence of types such as dependent lists which
become easy to use; many proofs get automated, making the life of the
user easier (developing the example of reverse for dependent lists in
the currently distributed version of Coq took us a day of work, and we
don't believe this can be shrinked to one hour); and proofs are much
smaller, some seemingly complex proofs becoming simple reflexivity
proofs.  We believe that the resulting style of proofs becomes much
closer to that of the working mathematician.

We have also explained the advantage of the approach insofar as it
allows to clearly separate computation from deduction, therefore
allowing for an incremental development of the kernel of the system.

So far, we have considered only decidable -equality-
theories. However, thanks to the decidability assumption, a decidable
non-equality theory can always be transformed into a decidable
equality theory over the type Bool of truth values equipped with its
usual operations.

There are still many directions to be investigated. A first is to
embed membership equational logic in \CIC along the lines of the
simpler embedding described here. A second is to consider the case of
dependent algebras instead of the simpler parametric algebras. This is
a much more difficult question, which requires using a stronger notion
of conversion in the main argument of an elimination, but would
further help us addressing other weaknesses of Coq.

Finally, we strongly believe that the use of decision procedures
outputing certificates when they succeed and explanations when they
fail will change our way of making formal, and enlarge the audience of
proof assistants.

{{\bf Acknowledgement.} We thank the Coq group for many useful
discussions and suggestions, and the referees for their useful
remarks.}


\end{document}